\documentclass[12pt,superscriptaddress,nofootinbib,showpacs,showkeys,preprintnumbers]{revtex4}

\usepackage{amsfonts}
\usepackage{mathrsfs}
\usepackage{amssymb}
\usepackage{amsmath}
\usepackage{graphicx,epsfig}
\usepackage{pstricks}
\usepackage{slashed}

\def\be{\begin{equation}}
\def\ee{\end{equation}}
\def\bea{\begin{eqnarray}}
\def\eea{\end{eqnarray}}
\def\bean{\begin{eqnarray*}}
\def\eean{\end{eqnarray*}}
\def\bary{\begin{array}}
\def\eary{\end{array}}
\def\bit{\begin{itemize}}
\def\eit{\end{itemize}}

\def\su5u1{SU(5) \times U(1)}
\def\fsu5u1{SU(5) \times U(1)'}
\def\so10{SO(10)}
\def\sq20{SO(10) \times SO(10)}

\def\bwt{\begin{widetext}}
\def\ewt{\end{widetext}}
\def\be{\begin{equation}}
\def\ee{\end{equation}}
\def\bea{\begin{eqnarray}}
\def\eea{\end{eqnarray}}
\def\bean{\begin{eqnarray*}}
\def\eean{\end{eqnarray*}}
\def\bary{\begin{array}}
\def\eary{\end{array}}
\def\bit{\begin{itemize}}
\def\eit{\end{itemize}}

\def\su5u1{SU(5) \times U(1)}
\def\fsu5u1{SU(5) \times U(1)'}
\def\so10{SO(10)}
\def\sq20{SO(10) \times SO(10)}

\usepackage{amssymb}

\begin{document}

\setlength{\parskip}{0.1cm}

\preprint{MIFPA-12-02,\, OSU-HEP-12-02}

\title{\Large Quark lepton unification in higher dimensions}

\author{Tianjun Li}
\email{tli@itp.ac.cn}
\affiliation{Key Laboratory of Frontiers in Theoretical Physics,
      Institute of Theoretical Physics, Chinese Academy of Sciences,
Beijing 100190, P. R. China }
\affiliation{George P. and Cynthia W. Mitchell Institute for
Fundamental Physics, Texas A$\&$M University, College Station, TX
77843, USA }
\author{Zeke Murdock} 

\affiliation{Marymount College, 30800 Palos Verdes Dr. East, 
Rancho Palos Verdes, CA 90275, USA} \altaffiliation{{\bf Present address} }\email{zekemurdock@gmail.com} 
\affiliation{Department of Physics and Oklahoma Center for High
Energy Physics, Oklahoma State University, Stillwater, OK
74078-3072, USA}
\author{S. Nandi}
\email{s.nandi@okstate.edu}
\affiliation{Department of Physics and Oklahoma Center for High
Energy Physics, Oklahoma State University, Stillwater, OK
74078-3072, USA}
\author{Santosh Kumar Rai}
\email{santosh.rai@okstate.edu}
\affiliation{Department of Physics and Oklahoma Center for High
Energy Physics, Oklahoma State University, Stillwater, OK
74078-3072, USA}

\begin{abstract}
The idea of unifying quarks and leptons in a gauge symmetry is very
appealing. However, such an unification gives rise to leptoquark
type gauge bosons for which current collider limits push their
masses well beyond the TeV scale. We present a model in the
framework of extra dimensions which breaks such quark-lepton
unification symmetry via compactification at the TeV scale. These
color triplet leptoquark gauge bosons, as well as the new quarks
present in the model, can be produced at the LHC with distinctive
final state signatures. These final state signals include high $p_T$
multi-jets and multi-leptons with missing energy, 
monojets with missing energy, as well as the heavy charged particles
passing through the detectors, which we also discuss briefly. The
model also has a neutral Standard Model singlet heavy lepton which is stable,
and can be a possible candidate for the dark matter.

\end{abstract}



\maketitle

\section{Introduction}
 
Unifying quarks and leptons in the same multiplet of a gauge symmetry group is very elegant, and 
answers naturally why the quarks have fractional charges while the leptons have integer charges. 
Such an unification has been achieved in the framework of partial unification in
$SU(4)_C \times SU(2)_L \times SU(2)_R$ ~\cite{Pati:1974yy}, or grand unification such as 
$SU(5)$~\cite{gg} or $SO(10)$~\cite{gfm}. The leptoquark gauge bosons present in such grand 
unification lead to proton decay which has not been observed so far, and pushes the mass scale of these 
leptoquark gauge bosons to $~ 10^{16}$ GeV. This is well beyond the reach of any present or future high 
energy collider. In the partial unification model, such as Pati-Salam type, since $B-L$ and the fermion 
number is conserved, there is no proton decay by the leptoquark gauge bosons. However, the leptoquark 
gauge boson here can cause rare meson decays, such as $K_L\rightarrow{\mu^+ e^{-}}$ and 
$K_L\rightarrow{\mu^- e^{+}}$ ~\cite{Valencia:1994cj}. The upper limit for the combined branching ratio 
for these two  rare modes is $4.7 \times 10^{-12}$~\cite{pdg}. This gives the mass of the exchanged 
leptoquark gauge boson to be $>2.3 \times {10^3}$ TeV. This is well above the current or 
future LHC reach. The question we address in this work is can we have a model in which the leptoquark 
gauge boson has a mass in the TeV scale which can be probed at the LHC?

One trick to achieve such a low scale for quark-lepton unification in the framework of $SU(4)_C$  is to pair 
the known quarks with some new leptons, and the known leptons with some new quarks. Then the 
leptoquark gauge bosons will couple known quarks with these new leptons, and the known leptons 
with new quarks. This will avoid the limits from the above flavor-violating rare meson decay limits, 
by choosing the mixing angles between the ordinary and new fermions small. But this will cause 
problems with the precision electroweak parameters, since these new fermions will be chiral and we 
are doubling chiral fermion sector of the Standard Model (SM).

However, such a scenario can be implemented by considering Pati-Salam type model in the framework of 
extra dimensions, say in five dimensions. Then upon orbifold compactification of the extra dimension, 
the zero modes of the new fermions can be projected out, leaving only the known quarks and leptons as 
the zero modes. The same orbifold compactification can be used to break the $SU(4)_C$ symmetry to 
$SU(3)_C$ and will give masses to the leptoquark gauge bosons at the compactification scale which can 
be chosen at a TeV. After compactification, in the four dimensional theory, the new fermions will have only 
Kaluza-Klein modes, and will be vector-like, thus avoiding any problem with the electroweak (EW) 
precision parameters. 
We will implement this scenario with the simplest Pati-Salam type model with the gauge symmetry 
$SU(4)_C \times U(1)_{I3R} \times SU(2)_L $ in five dimensions; and breaking the symmetry to 
$SU(3)_C \times SU(2)_L \times U(1)_{I3R} \times U(1)_{B-L}$ upon compactification to four dimensions. 
$U(1)_{I3R} \times U(1)_{B-L}$ will be broken down to $U(1)_Y$ using the usual Higgs mechanism. The 
leptoquark gauge bosons, as well as the new quarks (which will be the lightest KK excitations), with 
masses at the TeV scale can be produced at the LHC giving distinctive signals for new physics. We mention here that quark lepton unification using  higher dimension has been considered by Adibzadeh and P. Q. Hung \citep{hung}, however their gauge symemmtry is different from ours, and there main objective was obtaining naturally light Dirac neutrino and fermion localization.

The paper is organized as follows. In section II, we discuss a five dimensional non-supersymmetric formulation of the model.  In section III, we discuss the one loop radiative correction, and calculate the particle spectrum. The phenomenological implications of the model: the productions, decays, and the new physics signals are discussed in section IV.  In section V, we give a five dimensional supersymmetric version of the model. Section VI contains our conclusions.

\section{A non-supersymmetric 5D model: Model and Formalism}

Our gauge symmetry is $SU(4)_C \times SU(2)_L \times U(1)_{I3R}$ in
five dimensions. The $SU(4)_C$ symmetry unifying quarks and leptons
is broken down to $SU(3)_C \times U(1)_{B-L}$ by compactifying the
extra dimension, $y$ on a $S^1/(Z_2\times Z'_2)$ orbifold. (The breaking of gauge symmetry via orbifold 
compactification is very elegant and has been extensively used in the literature to build realistic 
models~\cite{kawa, GAFF, LHYN, AHJMR, LTJ1, Hall:2001tn, Haba:2001ci, Asaka:2001eh, LTJ2, 
Dermisek:2001hp, Li:2001tx, Gogoladze:2003ci, Gogoladze:2003yw, Li:2003ee}).
The remaining symmetry, $U(1)_{B-L}\times U(1)_{I3R}$ in our model in four dimensions (4D) is then 
broken down to the SM by using appropriate Higgs multiplets. All the particles, gauge bosons,
Higgs bosons, as well as the fermions, propagate in the bulk, similar to the
universal extra dimensions (UED)\cite{Appelquist:2000nn}. 

In the five-dimensional orbifold models the five-dimensional manifold is
factorized into the product of ordinary four-dimensional Minkowski
space-time $M^4$ and the orbifold $S^1/(Z_2\times Z'_2)$.  The
corresponding coordinates are $x^{\mu}$ ($\mu = 0, 1, 2, 3$) and
$y\equiv x^5$. The radius for the fifth dimension is $R$.  The
orbifold $S^1/(Z_2\times Z'_2)$ is obtained by $S^1$ moduloing the
equivalent class
\begin{eqnarray}
P:~~y \sim -y~,~P':~~ y' \sim -y'~,~\,
\end{eqnarray}
where $y'\equiv y-\pi R/2$.  There are two fixed points, $y=0$ and
$y=\pi R/2$.

The gauge fields for $SU(4)_C \times SU(2)_L \times U(1)_{I3R}$ in five dimensions are $A4^A_{M}$,
$A2^A_{M}$, and $A^R_M$, respectively, and $M = \mu, 5$. All belong to the adjoint 
representations of the corresponding gauge symmetry. For generical five-dimensional gauge 
fields $A_M^A$, we have four-dimensional gauge fields $A_{\mu}^A$ and real scalar fields
$A_5^A$. The fermion and Higgs representations in five dimensions have
\begin{eqnarray}
FL_L~=~{\mathbf{(4, 2, 0)}}~,~~ FU_R ~=~{\mathbf{(4, 1, 1/2)}}~,~~ FD_R
~=~{\mathbf{(4, 1, -1/2)}}~\,
\end{eqnarray}
and
\begin{eqnarray}
FL^{'}_L~=~{\mathbf{(4, 2, 0)}}~,~~ FU^{'}_R~=~{\mathbf{(4, 1,
1/2)}}~,~~ FD^{'}_R ~=~{\mathbf{(4, 1, -1/2)}}~,~\,
\end{eqnarray}
\begin{eqnarray}
H~=~{\mathbf{(1, 1, -1/2)}}~,~\,
\end{eqnarray}
where the numbers in the parentheses represent the quantum numbers
with respect to the gauge symmetry, $SU(4)_C \times SU(2)_L \times
U(1)_{I3R}$.

In the usual left and right handed notation, the particle contents
in $FL$, $FL^{'}$, $FU$, $FU^{'}$, $FD$,$FD^{'}$ are
\begin{align}
&FL_L = (q_L,l^{'}_L),& &FL_R = (q_R, l^{'}_R),&
&FL^{'}_L = (q^{'}_L, l_L),&  &FL^{'}_R = (q^{'}_R, l_R),&\nonumber \\
&FU_R = (U_R,N^{\prime}_R),& &FU_L = (U_L,N^{'}_L),&
&FU^{'}_R = (U^{'}_R,N_R),& &FU^{'}_L = (U^{'}_L,N_L),&\nonumber \\
&FD_R = (D_R,E^{\prime}_R),& &FD_L = (D_L,E^{'}_L),&
&FD^{'}_R = (D^{'}_R,E_R),& &FD^{'}_L = (D^{'}_L,E_L),&
\end{align}
where we neglect the family indices $i=1, ~2, ~3$.
These represent the fermions in one family. There are three such
families. Note that the fermion content in each family has been
quadrupled where $q_L, l_L, u_R, d_R, e_R$ and  $N_R$ represent the usual
fermions in a family (including a right-handed neutrino in each
family), and the primes represent the additional fermions with same
corresponding quantum numbers. Note that the leptoquark gauge
bosons connect ordinary quarks to the exotic (primed) leptons, and
ordinary leptons to exotic (primed) quarks. This will avoid the high
experimental bound on the masses of these leptoquarks, and will
allow us a low compactification scale for the $SU(4)_C \rightarrow
SU(3)_C \times U(1)_{B-L}$ breaking.

We now discuss the gauge symmetry breaking via orbifold compactification,
to break the $SU(4)_C$ gauge symmetry down to the $SU(3)_C\times
U(1)_{B-L}$ gauge symmetry. Under the $Z_2$ and $Z'_2$
parity operators $P$, and $P^{'}$, the vector fields transform as
\begin{align}
A_\mu(x^{\mu},y)&\to  A_\mu(x^{\mu},-y) = P A_\mu(x^{\mu}, y) P^{-1} , &
A_\mu(x^{\mu},y^{'})&\to A_\mu(x^{\mu},-y^{'}) = P^{'} A_\mu(x^{\mu}, y^{'})
P^{'-1} \nonumber \\
A_5(x^{\mu},y) &\to A_5(x^{\mu},-y) = - P A_5(x^{\mu}, y) P^{-1}, &
A_5(x^{\mu},y^{'}) &\to A_5(x^{\mu},-y^{'}) = - P^{'} A_5(x^{\mu}, y^{'}) P^{-1}
\end{align}
where $A_5$ is the $5$th component of the vector field.

The fermion multiplets belonging to the fundamental representation of
the gauge symmetry transform as
%
\begin{align}
            Z_2 :~~~~ FL_L(x^{\mu},y)&\to FL_L(x^{\mu}, -y)  = \eta  P FL_L(x^{\mu},y)~, \nonumber \\
Z_2^\prime : ~~~~ FL_L(x^{\mu},y^{'})&\to FL_L(x^{\mu}, -y^{'})  = \eta P^{'} FL_L(x^{\mu},y^{'})~, \nonumber \\
            Z_2 :~~~~ FL_R(x^{\mu},y) &\to FL_R(x^{\mu}, -y)  = -\eta P FL_R(x^{\mu},y)~, \nonumber \\
Z_2^\prime : ~~~~ FL_R(x^{\mu},y^{'})&\to FL_R(x^{\mu}, -y^{'}) = - \eta P^{'} FL_R(x^{\mu},y^{'})
\end{align}
and similarly for the other fields. In a short-hand notation, the
transformation properties of the fields are given by 
\begin{align*}
& FL_L : (\eta P, \eta P'), &  & FL_R : (- \eta P, -\eta P'), &
& FU_R:(\eta P, \eta P'),  &  & FU_L : (- \eta P, -\eta P'), \\
& FD_R:(\eta P, \eta P'),  &  & FD_L : (- \eta P, -\eta P'), &
& FL'_L:(\eta P,-\eta P'), &  & FL'_R : (- \eta P, \eta P'), \\
& FU'_R:(\eta P,-\eta P'), &  & FU'_L : (- \eta P, \eta P'), &
& FD'_R:(\eta P, -\eta P'), & & FD'_L : (- \eta P, \eta P') 
\end{align*}
where $\eta$ takes the value only $+1$.

To  project out the zero modes of the fifth component of gauge
fields, and the appropriate modes for the fermion fields, $FL$, and
$FL^{'}$ and others, we choose the following $4\times 4$ matrix
representations for the parity operators $P$ and $P'$
\begin{equation}
\label{eq:pppsu5} P={\rm diag}(+1, +1, +1, +1)~,~P'={\rm diag}(+1,+1, +1, -1)
 ~.~\,
\end{equation}
Under the $P'$ parity, the gauge generators $T^\alpha$ ($\alpha =1$,
$2$, ..., $15$) for $SU(4)_C$ are separated into two sets: $T^a$ are
the generators for the $SU(3)_C\times U(1)_{B-L}$ gauge group, and
$T^{\hat a}$ are the generators for the broken gauge group
\begin{equation}
P~T^a~P^{-1}= T^a ~,~ P~T^{\hat a}~P^{-1}= T^{\hat a} ~,~\,
\end{equation}
\begin{equation}
P'~T^a~P^{'-1}= T^a ~,~ P'~T^{\hat a}~P^{'-1}= - T^{\hat a} ~.~\,
\end{equation}
The zero modes of the $SU(4)_C/(SU(3)_C\times U(1)_{B-L})$ gauge
bosons are projected out, thus, the five-dimensional $SU(4)_C$ gauge
symmetry is broken down to the four-dimensional $SU(3)_C\times
U(1)_{B-L}$ gauge symmetry. For the $SU(2)_L$ gauge symmetry, we
choose
\begin{equation}
P~=~P'~=~{\rm diag}(+1, +1)
 ~,~\,
\end{equation}
and for the $U(1)_{R}$ gauge symmetry, we choose $P=P'=1$. Thus, the
$SU(2)_L \times U(1)_{I3R}$ remains unbroken after orbifold
compactification.

Denoting the generical fields  $\phi$ with parities ($P$,
$P'$)=($\pm, \pm$) by $\phi_{\pm \pm}$, we obtain the  KK mode
expansions
\begin{eqnarray}
  \phi_{++} (x^\mu, y) &=&
      \sum_{n=0}^{\infty} \frac{1}{\sqrt{2^{\delta_{n,0}} \pi R}}
      \phi^{(2n)}_{++}(x^\mu) \cos{2ny \over R}~,~\,
\end{eqnarray}
\begin{eqnarray}
  \phi_{+-} (x^\mu, y) &=&
      \sum_{n=0}^{\infty} \frac{1}{\sqrt{\pi R}}
      \phi^{(2n+1)}_{+-}(x^\mu) \cos{(2n+1)y \over R}~,~\,
\end{eqnarray}
\begin{eqnarray}
  \phi_{-+} (x^\mu, y) &=&
      \sum_{n=0}^{\infty} \frac{1}{\sqrt{\pi R}} \,
      \phi^{(2n+1)}_{-+}(x^\mu) \sin{(2n+1)y \over R}~,~\,
\end{eqnarray}
\begin{eqnarray}
  \phi_{--} (x^\mu, y) &=&
      \sum_{n=0}^{\infty} \frac{1}{\sqrt{\pi R}}
      \phi^{(2n+2)}_{--}(x^\mu) \sin{(2n+2)y \over R}~,~\,
\end{eqnarray}
where  $n$ is a non-negative integer. The four-dimensional fields
$\phi^{(2n)}_{++}$, $\phi^{(2n+1)}_{+-}$, $\phi^{(2n+1)}_{-+}$ and
$\phi^{(2n+2)}_{--}$ acquire masses $2n/R$, $(2n+1)/R$, $(2n+1)/R$
and $(2n+2)/R$ upon the compactification. Zero modes are contained
only in $\phi_{++}$ fields.
Moreover, only $\phi_{++}$ and $\phi_{+-}$ fields have non-zero
values at $y=0$, and only $\phi_{++}$ and $\phi_{-+}$ fields have
non-zero values at $y=\pi R/2$.

\begin{table}[h!]
\begin{center}
\begin{tabular}{|c|c|c|}
\hline $(P,P')$ & ${\rm Field}$ & ${\rm Mass}$ \\ \hline
$(+,+)$ & $A4_{\mu}^a$, $A2_{\mu}^A$, $A^R_{\mu}$, $q_L$, $U_R$, $D_R$, $l_L$,
$N_R$, $E_R$, $H$  & ${{2n}\over R}$ \\ \hline
$(+,-)$ &  $A4_{\mu}^{\hat{a}}$, $q'_L$, $U^{\prime}_R$, $D^{\prime}_R$, $l^{'}_L$,
$N^{\prime}_R$, $E^{\prime}_R$ & ${{2n+1}\over R}$  \\ \hline
$(-,+)$ &  $A4^{\hat{a}}_5$, $q^{'}_R$, $U^{'}_L$, $D^{\prime }_L$,
$l^{'}_R$, $N^{\prime}_L$, $E^{\prime}_L$ & ${{2n+1}\over R}$  \\ \hline
$(-,-)$ & ~$A4_5^a$, $A2_{5}^A$, $A^{R}_{5}$, $q_R$, $U_L$, $D_L$,
$L_R$, $N_L$, $E_L$, ~
  & ${{2n+2}\over R}$ \\ \hline
\end{tabular}
\end{center}
\caption{Parity assignments and masses ($n\ge 0$) for  the bulk
fields. } \label{Spectra-I}
\end{table}
The particle spectra and their $(P,~P')$ parity are given in
Table~\ref{Spectra-I}. Note that for each KK excitations, the left
and right handed parts of each four-dimensional field combine to form
a massive Dirac spinor.

The $SU(4)_C$ symmetry in our model is broken down to $SU(3)_C \times
U(1)_{B-L}$ by the orbifold compactification $S^1 / Z^{'}_2$.
To break the $ U(1)_{B-L} \times U(1)_{I3R}$  gauge symmetry down
to the $U(1)_Y$ gauge symmetry and give the Majorana masses
to the right-handed neutrinos, we introduce a SM singlet
Higgs field $S$ which is localized on the 3-brane at
$y=\pi R/2$ and has a vacuum expectation
value (VEV) at the TeV scale. The quantum number of $S$ under
$SU(3)_C\times SU(2)_L \times U(1)_{B-L} \times U(1)_{I3R}$ gauge symmetry
is $(\mathbf{1}, \mathbf{1}, \mathbf{-1},  \mathbf{1})$.
Finally, the $SU(2)_L \times U(1)_Y$ gauge symmetry breaking to $U(1)_{em}$
is achieved by using the usual standard model doublet Higgs
$H$ with VEV
at the electroweak scale. The complete SM fermion Yukawa couplings
are
\begin{eqnarray}
-{\cal L} &=& \left( \lambda^u_{ij} (FU^j_R)^c FL^i_L \widetilde{H}
+\lambda^d_{ij} (FD^j_R)^c FL^i_L H +
\lambda^{\prime u}_{ij} (FU^{\prime j}_R)^c FL^{\prime i}_L \widetilde{H}
+ \lambda^{\prime d}_{ij} (FD^{\prime j}_R)^c FL^{\prime i}_L H \right)
\nonumber \\ &&+  \left( h^u_{ij} (FU^j_R)^c FL^i_L \widetilde{H}
+h^d_{ij} (FD^j_R)^c FL^i_L H +
h^{\prime u}_{ij} (FU^{\prime j}_R)^c FL^{\prime i}_L \widetilde{H}
+ h^{\prime d}_{ij} (FD^{\prime j}_R)^c FL^{\prime i}_L H  \right)
\delta (y) \nonumber \\ &&
+ \left( y^u_{ij} (U_R^j)^c q^i \widetilde{H}
+y^d_{ij} (D_R^j)^c q^i  H +
y^{\nu}_{ij} (N_R^j)^c l^i_L \widetilde{H}
+ y^{e}_{ij}  (E_R^j)^c l^i_L H +y^N_{ij} S^{\dagger} N_R^i N_R^j  \right)
\nonumber \\ &&
\times \delta (y-\pi R/2) ~,~\,
\end{eqnarray}
where $\widetilde{H} = i\sigma_2 H^*$ and $\sigma_2$
is the second Pauli matrix.

We define the $U(1)_{B-L}$ generator in $SU(4)_C$ as follows
\begin{eqnarray}
T_{\rm U(1)_{B-L}}={\rm diag} \left({1\over 6}, {1\over 6}, {1\over
6},
 -{1\over 2} \right)~.~\,
\label{u1B-L}
\end{eqnarray}
Thus, we obtain the $U(1)_{B-L}$ gauge coupling $g_{B-L}$ and
$SU(3)_C$ gauge coupling $g_3$ at the compactification scale in
terms of $SU(4)_C$ gauge coupling $g_{4C}$
\begin{eqnarray}
g_{B-L}~=~{\sqrt {3\over 2}} ~g_{4C}~,~~g_3=g_{4C}~.~\,
\end{eqnarray}

We denote the $U(1)_Y$ gauge field as $B_{\mu}$, and the orthogonal
massive $U(1)$ gauge field as $Z'_{\mu}$. Thus, we obtain
\begin{eqnarray}
\left(
\begin{array}{c}
B_\mu \\
Z'_\mu
\end{array} \right)=
\left(
\begin{array}{cc}
\cos\theta & \sin\theta \\
-\sin\theta & \cos\theta
\end{array}
\right) \left(
\begin{array}{c}
A_{\mu}^R \\
A_{\mu}^{B-L}
\end{array} \right)
~,~\,
\end{eqnarray}
where
\begin{eqnarray}
\sin\theta \equiv {{g_R}\over\displaystyle {\sqrt {g_{B-L}^2
+g_R^2}}} ~,~ \cos\theta \equiv {{g_{B-L}}\over\displaystyle {\sqrt
{g_{B-L}^2 +g_R^2}}} ~.~\,
\end{eqnarray}
And the $Z'_{\mu}$ mass is
\begin{eqnarray}
M_{Z'}~=~ {\sqrt {2 (g^2_{B-L}+g_R^2)v^2}}  ~.~\,
\end{eqnarray}
Moreover, we obtain the $U(1)_Y$ gauge coupling as follows
\begin{eqnarray}
g_Y^2= {{g_{B-L}^2 g_R^2 }\over\displaystyle{g_{B-L}^2 +g_R^2}}
~.~\,
\end{eqnarray}
The covariant derivative for $B_{\mu}$ and $Z'_{\mu}$ is
\begin{eqnarray}
D_{\mu} & \equiv & \partial_{\mu} - i Y_{B-L} g_{B-L} A_{\mu}^{B-L}
-i Y_R g_R A_{\mu}^R + \cdots \nonumber\\
&=&
\partial_{\mu} - i g_Y \left( Y B_{\mu} + Y_{Z'} Z'_{\mu} \right) + \cdots ~,~ \,
\end{eqnarray}
where $Y$ is the hypercharge, and
\begin{eqnarray}
Y~=~ Y_{B-L} + Y_R ~,~~~Y_{Z'}~=~  Y_{B-L} \cot\theta - Y_R
\tan\theta ~.~ \,
\end{eqnarray}

When the $SU(4)_C$ gauge symmetry is broken down to the
$SU(3)_C\times U(1)_{B-L}$ gauge symmetry, we can arrange the broken
gauge fields into the vector fields $A'_{\mu}$ and
$\overline{A'}_{\mu}$, whose quantum numbers under $SU(3)_C\times
SU(2)_L \times U(1)_Y$ are $(\mathbf{3}, \mathbf{1}, \mathbf{2/3})$
and $(\mathbf{\overline{3}}, \mathbf{1}, \mathbf{-2/3})$. Their
interactions with the matter fields can be obtained from
 the following covariant derivative
\begin{eqnarray}
D_{\mu} & \equiv & \partial_{\mu} - i {{g_{4C}}\over {\sqrt 2}}
\left(
\begin{array}{cc}
G^a_{\mu} \lambda^a + A^{B-L}_{\mu} I_{3\times 3} /6 & A'_{\mu} \\
\overline{A'}_{\mu} & -A^{B-L}_{\mu}/2
\end{array}
\right) + \cdots ~,~\,
\end{eqnarray}
where $G^a_{\mu}$ are the gluon fields, $\lambda^a$ are Gell-Mann
matrices, and $ I_{3\times 3}$ is the $3\times 3$ identity matrix.

\section{One loop radiative correction and particle spectrum}

Note that in our model, first KK excitations of the the particles
belonging to $(++)$ and $(--)$ have masses $2 / R$, while those
belonging to $(+-)$ and $(-+)$ have masses $1 / R$. Thus at the
tree level, all of the $(+-)$ and $(-+)$ are degenerate. However,
radiative corrections  will split these masses. The candidate for
the dark matter, the decay pattern of these particles and the
associated collider phenomenology will depend crucially on these
radiative splittings.

The radiative corrections to the KK particle masses have not been
calculated for the models on $S^1/(Z_2\times Z'_2)$. To have some
idea about radiative corrections, we consider the  $SU(4)_C \times
SU(2)_L \times U(1)_{B-L}$ model on $S^1/Z_2$, {\it i.e.}, we do not
break the $SU(4)_C $ gauge symmetry via $Z_2$ orbifold projections.
Using the generic formulae for radiative corrections
 given in Ref.~\cite{Cheng:2002iz}, we obtain
the bulk radiative corrections to the KK particle masses
\begin{eqnarray}
\delta\, (m_{A_{n}^{R}}^2) &=&  -\frac{47}{2}\,
\frac{g_R^2\,\zeta(3)}{16\pi^4} \left(\frac{1}{R}\right)^2,
\nonumber \\
\delta\, (m_{W_n}^2) &=&
-\frac{17}{2}\,\frac{g_2^2\,\zeta(3)}{16\pi^4}
\left(\frac{1}{R}\right)^2,
\nonumber \\
\delta\, (m_{A4_n}^2) &=& -6 \, \frac{g_{4C}^2\,\zeta(3)}{16\pi^4}
\left(\frac{1}{R}\right)^2,
\nonumber \\
\delta( m_{F_{Ln}}) &=& \delta( m_{FU_Rn})~=~\delta(
m_{FD_Rn})~=~\delta(m_{H_n}^2) ~=~0~,~\, \label{Bulk-MC}
\end{eqnarray}
where $\zeta(3)$ is about 1.202.

The boundary terms receive divergent contributions that require
counter terms. The finite parts of these counter terms are
undetermined and remain as free parameters of the theory. Assuming
that the boundary kinetic terms vanish at the cutoff scale $\Lambda$
and calculate their renormalization to the lower energy scale $\mu$,
we obtain the radiative corrections from the boundary terms
\begin{eqnarray}
\bar{\delta}\, (m_{A^R_n}^2) &=& m_n^2\, \left(-\frac{1}{6}\right)\,
\frac{g_R^2}{16\pi^2}
\ln\frac{\Lambda^2}{\mu^2}, \nonumber \\
\bar{\delta}\, (m_{W_n}^2) &=& m_n^2\, \frac{15}{2}\,
\frac{g_2^2}{16\pi^2}
\ln\frac{\Lambda^2}{\mu^2}, \nonumber \\
\bar{\delta}\, (m_{A4_n}^2) &=& m_n^2 \,\frac{46}{3}\,
\frac{g_{4C}^2}{16\pi^2}
\ln\frac{\Lambda^2}{\mu^2}, \nonumber \\
\bar{\delta}\, m_{FLn} &=& m_n\,\left(\frac{135}{32}\,
\frac{g_{4C}^2}{16\pi^2} + \frac{27}{16}\,\frac{g_2^2}{16\pi^2}
\right) \,\ln\frac{\Lambda^2}{\mu^2},
\nonumber \\
\bar{\delta}\, m_{FUn} &=& \bar{\delta}\, m_{FDn}~=~
m_n\,\left(\frac{135}{32}\, \frac{g_{4C}^2}{16\pi^2} +
\frac{9}{16}\,\frac{g_R^2}{16\pi^2} \right)
\,\ln\frac{\Lambda^2}{\mu^2},
\nonumber \\
\bar{\delta}\, (m_{H_n}^2) &=& m_n^2 \, \left(\frac{3}{2}\, g_2^2 +
\frac{3}{4}\, g_R^{2} - \lambda_H \right)
\frac{1}{16\pi^2}\ln\frac{\Lambda^2}{\mu^2} + \overline{m}_H^2 ~,~\,
\label{Boundary-MC}
\end{eqnarray}
where  $\lambda_H$ is the Higgs quartic coupling (${\cal L}\supset
-(\lambda_H/2) (H^\dagger H)^2$), and $\overline{m}_H^2$ is the
boundary Higgs mass term. The renormalization scale $\mu$ should be
taken to be approximately the mass of the corresponding KK mode. For
simplicity, we neglect the Yukawa coupling corrections here since
their corrections are about one or two percents even for order one
Yukawa couplings. Thus, comparing to the boundary corrections, the
bulk corrections to the $SU(4)_C\times SU(2)_L$ gauge bosons' masses
are indeed very small since in addition to the coefficients, there
is another $\pi^2$ suppressions. However, for $U(1)_R$ gauge boson,
the bulk corrections are larger than the boundary corrections. In
addition, the boundary corrections to $SU(4)_C$ gauge bosons' masses
are much larger than those to the $FL$, $\overline{FU}$ and
$\overline{FD}$.
\begin{figure}[!b]
\includegraphics[width=3.2in]{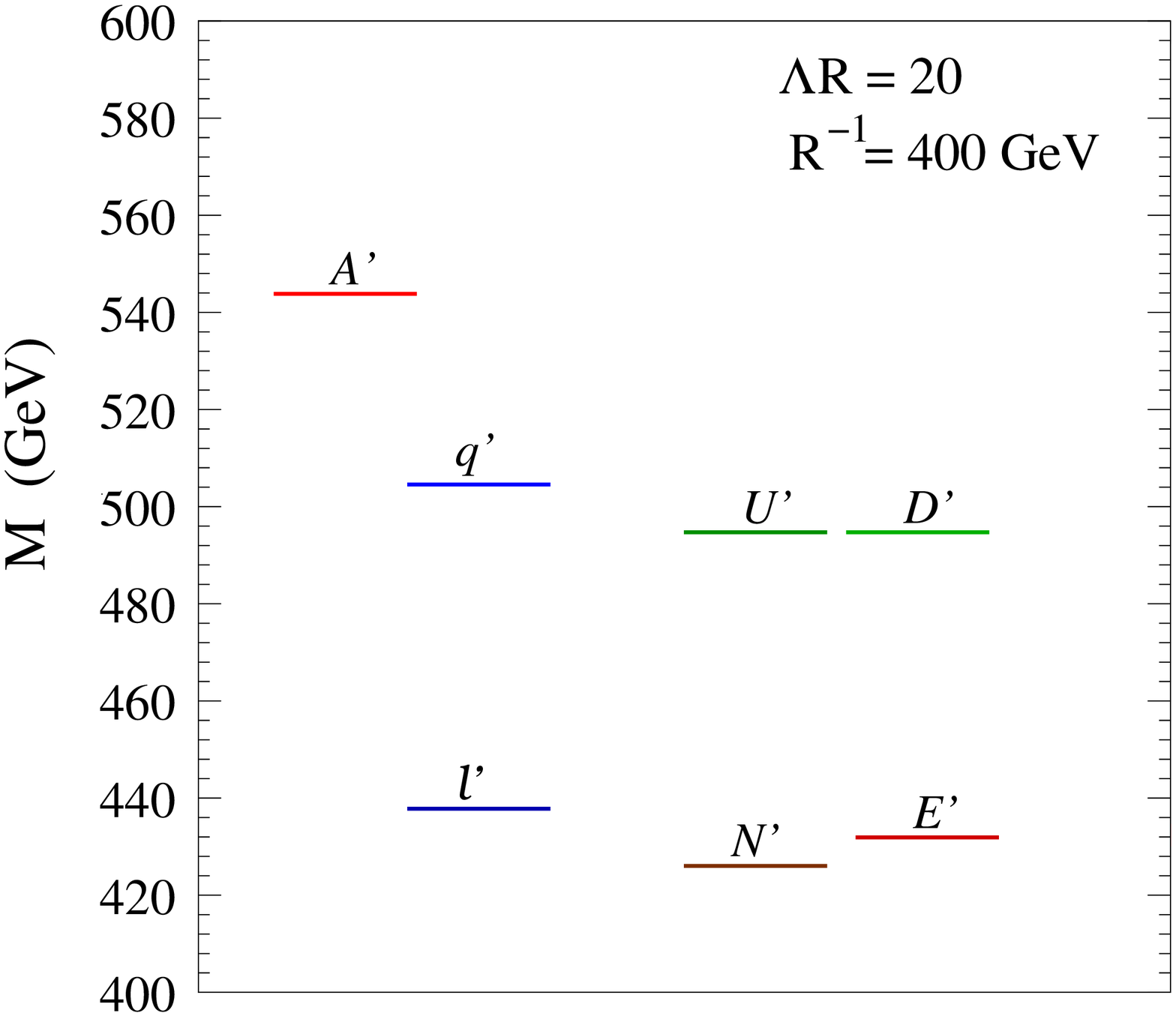}
\includegraphics[width=3.2in]{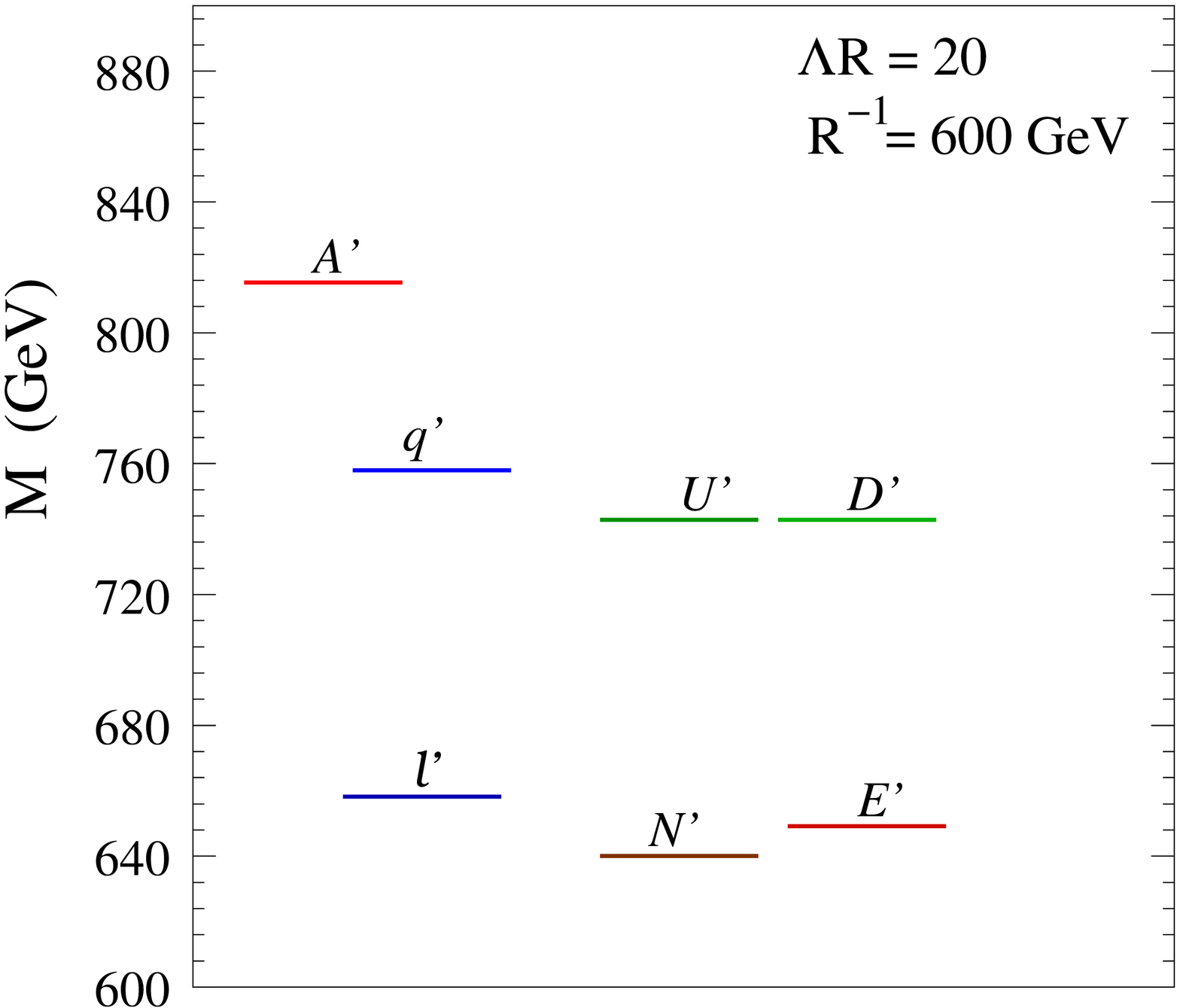}
\caption{\sl The particle mass spectrum of the $(+-)$ states for two values of the compactification
scale $R$ after including the mass splittings.}
\label{fig:mass_spectrum}
\end{figure}

The gauge symmetry $SU(4)_C$ is broken down to $SU(3)_C \times
U(1)_{B-L}$ by the orbifold compactification $S^1 / Z^{'}_2$. The
$SU(3)$ color interaction will split the masses among the members of
the $(+-)$ modes. We estimate these splittings as
\begin{eqnarray}
\bar{\delta}\, m_{q^{'}_{Ln}}- \bar{\delta}\, m_{l^{'}_{Ln}}&=&
m_n\,3 \frac{g_{3C}^2}{16\pi^2}\ln\frac{\Lambda^2}{\mu^2},
\end{eqnarray}
\begin{eqnarray}
\bar{\delta}\, m_{u^{'}_{Rn}}- \bar{\delta}\, m_{N^{'}_{Rn}}&=&
m_n\, \frac{3 g_{3C}^2 + g'^2}{16\pi^2}\ln\frac{\Lambda^2}{\mu^2},
\end{eqnarray}
\begin{eqnarray}
\bar{\delta}\, m_{d^{'}_{Rn}}- \bar{\delta}\, m_{E^{'}_{Rn}}&=&
m_n\, \frac{3 g_{3C}^2 - 2 g'^2}{16\pi^2}\ln\frac{\Lambda^2}{\mu^2},
\end{eqnarray}
where
\begin{align}
\bar{\delta}\, m_{q^{'}_{Ln}}= \delta( m_{FLn}), &&
\bar{\delta}\, m_{u^{'}_{Rn}}= \delta( m_{FURn}),&&
\bar{\delta}\, m_{d^{'}_{Rn}}= \delta( m_{FDRn}),
\end{align}
the expressions for which are given before. Using the above
equations, the spectrum of the first KK excitations of the particles
can be calculated. Two parameters in the model are $1 /R$ and
$\Lambda$. In Fig. \ref{fig:mass_spectrum}, we have shown the particle spectrum  for $ 1 / R
= 400$ GeV, $\Lambda R = 20$, and for $1/R = 600 $ GeV, and $\Lambda R = 20$. We also list the
explicit values of the masses in Table \ref{tab:mass_spectrum}.

\begin{figure}[!htb]
\includegraphics[width=3.2in]{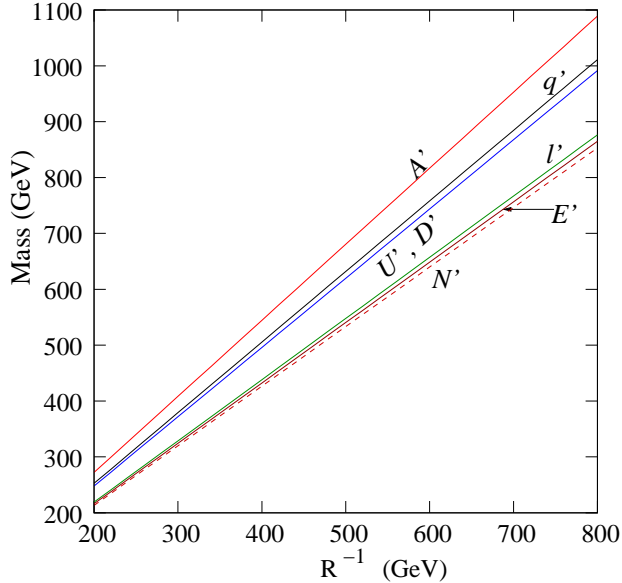}
\caption{\sl The particle mass spectrum as a function of the compactification scale $R^{-1}$ after
including the mass splittings.}
\label{fig:mass_spectrum2}
\end{figure}

After calculating the mass splittings, the heaviest state in the particle spectrum for the
first KK excited states with parity properties $(+-)$ turns out to be the fractionally
charged colored vector boson $A'$. The colored fermion states are much heavier than the
lepton excitations and the lowest lying particles in the mass spectrum are the neutral
$SU(2)_L$ and color singlet excitation ($N^{\prime}$) of the right-handed neutrino, which
we call as the lightest exotic particle (LXP).
In Fig. \ref{fig:mass_spectrum2} we show the mass of the KK excitations as a function of
the compactification scale $R^{-1}$. The larger values of the compactification scale lead
to larger mass splittings between the different particles.

Just like any other model with some discrete parity symmetry preventing the decay of the
lightest particle into only SM particles, the LXP in our model ($N^\prime$) too
is stable and a candidate for cold dark matter.
\begin{center}
\begin{table}[t!]
\begin{tabular}{|l||c|c|c|c|c|c|c|}\hline
Particles  &  $A'$  &  $q^{'i}$  &  $U^{'i}$ &  $D^{'i}$ &  $l^{'i}$ &  $E^{'i}$ &  $N^{'i}$ \\ \hline
$R^{-1}=400$ &  544.5 & 505.5 & 495.8 & 495.8 & 438.2 & 432.4 & 426.6    \\ \hline
$R^{-1}=600$ &  816.7 & 758.2 & 743.7 & 743.7 & 657.3 & 648.6 & 639.9    \\ \hline
\end{tabular}
\caption{\sl The masses (in GeV) for the KK excitations of the $(+-)$ states for two values of the
compactification scale $R$ where $\Lambda R=20$.}\label{tab:mass_spectrum}
\end{table}
\end{center}

\section{Phenomenological Implications: Decay modes and collider signals}

\begin{figure}[!t]
\includegraphics[width=3.2in]{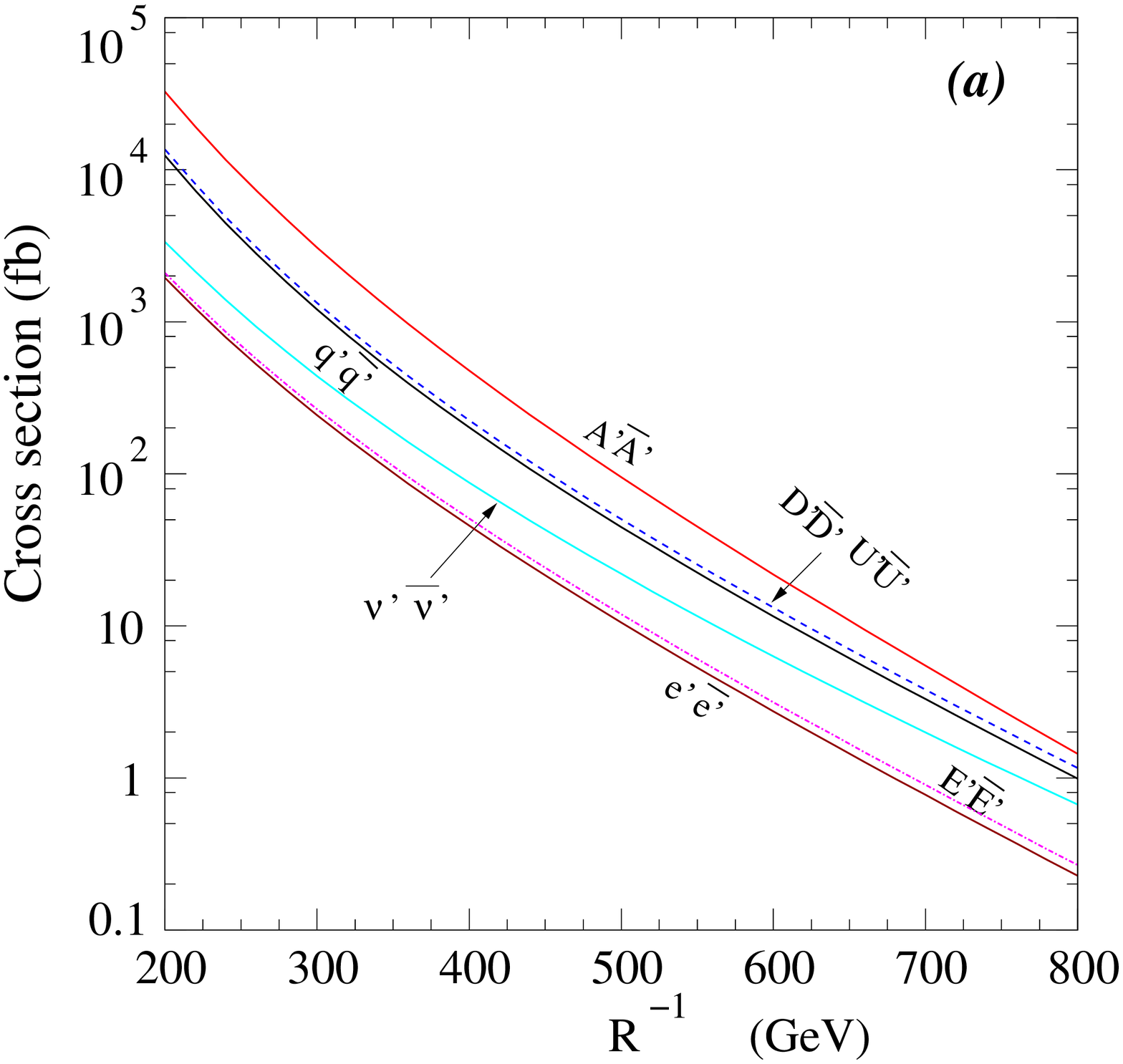}
\includegraphics[width=3.2in]{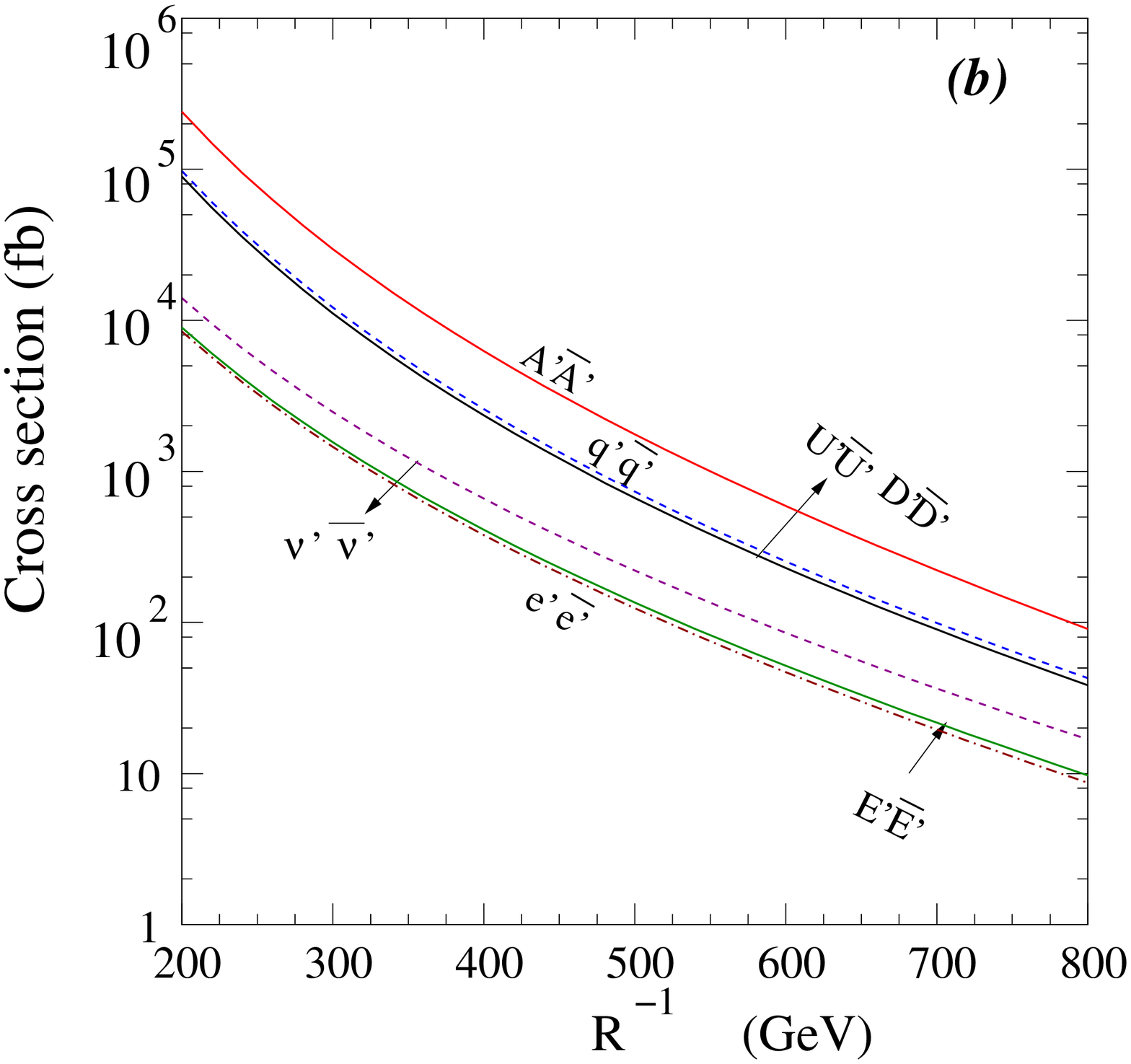}
\caption{\sl Illustrating the pair production cross sections for the first KK excitations as
a function of the compactification scale $R^{-1}$ at LHC for the center-of-mass energies of
(a) $\sqrt{s}=7$ TeV and (b) $\sqrt{s}=14$ TeV.}
\label{fig:cross_sections}
\end{figure}

By construction one expects the phenomenology of this model to have features present in the UED type models (for example, see \cite{uedmodels}, where one produces the KK excitations of the SM
particles and studies the multi-lepton and multi-jet final states accompanied with a large
missing energy carried away by the lightest KK particle. However, the motivation of our model 
is completely different, namely, the quark lepton unification. Thus our model has leptoquark gauge bosons, 
as well as new fermions not present in the UED type models.  Also, in our model, the first KK excitations of the 
SM particles are much heavier with the mass scale of $2/R$, instead of $1/R$, whereas the KK excitations of 
the new particles ($q', l'$) have mass scale of $1/R$. These particles are not present in the UED type models. The
interesting signal in our model arises from the production of the KK excitations of the
new strongly interacting exotic fermions ($q^{\prime},U^{\prime},D^{\prime}$) as well as
the KK excitations of the colored leptoquark gauge boson ($A^\prime$). All these modes have large pair 
production cross sections as their production proceeds through strong interactions. In 
Fig. \ref{fig:cross_sections} we plot the pair production cross sections for the strongly interacting 
exotics as well as the leptonic modes at LHC for two different center of mass energies $7$ and $14$ TeV.
It is worth noting that the fractionally charged vector boson $A^\prime$ has the largest
production cross section at the LHC and it couples directly to the gluon with the
interaction vertices ($G~A^\prime~A^\prime$) and ($G~G~A^\prime~A^\prime$).  Also, it
mediates the decay of the exotic fermions. The decay of these exotics determine the
phenomenological implications of this model at collider experiments. To highlight the
interesting signal this model has at colliders, we look at all the possible decay modes
that these exotics have for the two values of $R^{-1}=400$ GeV and $R^{-1}=600$ GeV. The
dominant decay channels along with their branching ratios are collected in Table
\ref{tab:dk_Xprime}, which we have calculated using {\tt CalcHEP} \cite{Pukhov:2004ca}.
\begin{table}[t!]
\begin{center}
\begin{tabular}{|l|c|c|c|}\hline
Decay Modes  &  $R^{-1}=400$ GeV & $R^{-1}=600$ GeV & i \\ \hline
\hline BR ($A' \to \bar{e}^i D^{'i}_R$) & 2.03 \% & 1.99 \% & 1,2,3 \\
\hline BR ($A' \to \nu_L^i u^{'i}_L$) & 1.34 \% & 1.31 \% & 1,2,3 \\
\hline BR ($A' \to \bar{e}^i d^{'i}_L$) &1.34 \% & 1.31 \% & 1,2,3 \\
\hline BR ($A' \to d^i \bar{E}^{'i}_R$) & 8.81 \% & 8.64 \% & 1,2,3 \\
\hline BR ($A' \to d^i \bar{e}^{'i}_L$) & 8.06 \% & 7.91 \% & 1,2,3  \\
\hline BR ($A' \to u^i \nu^{'i}_L$) & 8.06 \% & 7.91 \% & 1,2 \\
\hline BR ($A' \to u^i N^{'i}_R$) & 9.57 \% & 9.38 \% & 1,2 \\
\hline BR ($A' \to t N^{'3}_R$) & -- & 1.94 \% & \\ \hline
\hline BR ($ q^{'1}_L\to \nu_L^i d^i \bar{E}^{'i}_R$) & 10.8 \% & 10.8 \% & 1,2,3 \\
\hline BR ($ q^{'1}_L\to \nu_L^i d^i \bar{e}^{'i}_L$) & 7.7 \% & 7.7 \% & 1,2,3 \\
\hline BR ($ q^{'1}_L\to \nu_L^i u^i \nu^{'i}_L$) & 7.7 \% & 7.7 \% & 1,2 \\
\hline BR ($ q^{'1}_L\to \nu_L^i u^i N^{'i}_R$) & 14.6 \% & 14.6 \% & 1,2 \\ \hline
\hline BR ($ D^{'1}_R\to e d^i \bar{E}^{'i}_R$) & 10.9 \% & 10.9 \% & 1,2,3 \\
\hline BR ($ D^{'1}_R\to e d^i \bar{e}^{'i}_L$) & 7.3 \% & 7.3 \% & 1,2,3 \\
\hline BR ($ D^{'1}_R\to e u^i \nu^{'i}_L$) & 7.3 \% & 7.3 \% & 1,2,3 \\
\hline BR ($ D^{'1}_R\to e u^i N^{'i}_R$) & 15.6 \% & 15.6 \% & 1,2 \\ \hline
\hline BR ($ \nu^{'1}_L\to u \bar{u}^i N^{'i}_R$) & 50 \% & 49 \% & 1,2 \\ \hline
\hline BR ($ e^{'1}_L\to \bar{u} d N^{'1}_R$) & 49.5 \% & 48.6 \% &  \\
\hline BR ($ e^{'1}_L\to \bar{c} d N^{'2}_R$) & 47.1 \% & 47.6 \% &  \\ \hline
\hline BR ($ E^{'1}_R\to \bar{u} d N^{'1}_R$) & 55.0 \% & 52.3 \% &  \\
\hline BR ($ E^{'1}_R\to \bar{c} d N^{'2}_R$) & 45.0 \% & 47.7 \% &  \\ \hline
\end{tabular}
\caption{\sl The dominant decay modes and the respective branching ratios for the first KK excitation
of the new exotic particles for two values of the compactification scale.} \label{tab:dk_Xprime}
\end{center}
\end{table}

From Table \ref{tab:dk_Xprime}, we note that the dominant decay channel for the leptoquark gauge boson, $A'$  is to a light jet and
the lightest KK particle, $N'$ (around 10\%). This would mean that the dominant signal in this
case becomes 2 jets with large missing energy. The other dominant decay
channels are $A' \to d^i \bar{E}^{'i}_R;~ d^i \bar{e}^{'i}_L$ and $A' \to u^i \nu^{'i}_L$.
However all the subsequent decay of the above exotics lead to multi-jet final states with
a large amount of missing energy carried away by the LXP, $N'$. It is
worth noting that all decays other than that of $A^\prime$ are three-body decays since
they all decay via $A'$ exchange. Since the mass difference between the
exotic quarks and the exotic leptons is quite large, the jets and leptons coming from the
exotic quark decays would be much harder as compared to the jets that come from the exotic
leptons. Thus even though one expects a high jet multiplicity in the signal, the
sub-leading jets coming from decays of $e',\nu'$ and $E'$ would be quite soft.  Although a
large jet multiplicity is nothing new at a hadron machine like the LHC, we must note that
all the new exotics would finally decay to the LXP ($N'$). Thus,
along with a large jet multiplicity in the signal for the new exotics one needs to trigger
on the final configuration demanding a large amount of missing energy which would very
clearly be able to suppress the SM background without affecting the signal too much. One
can also have multi-lepton final states but with lower signal events because of the
smaller branching fractions as shown in Table \ref{tab:dk_Xprime}. We must however
note that the missing transverse energy ($\slashed{E}_T$) for our signal crucially depends
on the mass splittings ($\Delta M$) between the LXP ($N^\prime$) and the heavy
exotic it comes from. A small mass splitting would lead to smaller $\slashed{E}_T$ even
for a heavy LXP as it would carry away very small kinetic energy. The exact behavior of the
signal would require a detail simulation of the events in the hadronic collider environment
which we leave for future studies.
\begin{center}
\begin{table}
    \begin{tabular}{|l|c|c|}
    \cline{2-3}
        \multicolumn{1}{c|}{}& $R^{-1}=400$ GeV & $R^{-1}=600$ GeV\\ \hline
        \bf{Signal} & $\sigma \times BR$ (fb) & $\sigma \times BR$ (fb)\\\hline \hline
        2 hard jets and $\slashed{E}_T$ & 8330 & 787\\  \hline
        4 hard jets and $\slashed{E}_T$ & 611 & 53.6\\\hline
        1 lepton, 3 hard jets and $\slashed{E}_T$ & 1890 & 102\\    \hline
        2 leptons, 2 hard jets and $\slashed{E}_T$ & 5020 & 393\\   \hline
        2 leptons, 4 hard jets and $\slashed{E}_T$ & 30.0 & 2.75\\\hline
        4 leptons, 2 hard jets and $\slashed{E}_T$ & 30.0 & 2.75\\ \hline
    Monojet ($p_T^j>20$ GeV) and $\slashed{E}_T$ & 2000 & 215\\ \hline
    \end{tabular}
\caption{\sl Illustrating the $\sigma\times BR$ for the various final states obtained for the
signal from the production and decay of the exotics at LHC with center-of-mass energy 
$\sqrt{s}=14$\ TeV. The leptons considered in the final states are either $e$ or $\mu$.}
\label{tab:collider}
\end{table}
\end{center}
The various important collider signatures are summarized in Table \ref{tab:collider}.  We also
give the $\sigma \times \mathrm{branching ~ratio (BR)}$ for the different final state topologies
which highlight the important signatures of our model.  All the
leptons are either $e$ or $\mu$ while the $\tau$'s have been considered as jets.  No cuts 
have been made on the signal although we expect the basic acceptance cuts to reduce the 
signal by  only 10\%-20\%. Note that the numbers shown are for a center-of-mass energy of
$\sqrt{s}=14$ TeV at LHC. The corresponding numbers for the current 
center-of-mass energy of $\sqrt{s}=7$ TeV at which the LHC is running can be obtained by a simple scaling of the $\sigma \times ~BR$ given in Table \ref{tab:collider}. The approximate scaling factors 
for $R^{-1}=400$ GeV and $R^{-1}=600$ GeV are $0.1$ and $0.05$ respectively. This implies that
some of the final states listed might be difficult to observe in the current run of LHC but will have 
much higher and observable rates when the machine runs at a higher center-of-mass energy. 

The signature common to all signals is a substantial amount of $\slashed{E}_T$ because of the
undetected  $N'$ and neutrinos present in the final states, coming from the decay chains shown
in Table \ref{tab:dk_Xprime}. The leptonic signals primarily come from the decay of the
$A^\prime$ and the $D_R^\prime$ exotics and give relatively large signal rates for the final
state configurations of $1\ell+3j+\slashed{E}_T$ and $2\ell+2j+\slashed{E}_T$. The 4-lepton
signature with at least 2 hard jets also looks very promising because of the very small SM
background expected for such a signal. In addition, the model has large cross section for the 
monojet plus large $\slashed{E}_T$ signal. Each jet is considered hard if it's $p_T > 40$ GeV. 
We discuss the individual signals and the possible SM backgrounds below.
\subsubsection*{\bf{$n$ hard jets and $\slashed{E}_T$}}
Hadronic signals with missing $\slashed{E}_T$ appear with very large cross section in our model. 
However such signals are hard to analyze at a hadronic machine, unless the associated missing 
transverse energy is very big. With the mass splittings in our model not very big, we do not expect a 
very extravagant value for the  $\slashed{E}_T$. There has been recent interest in looking for new 
physics signals in dijet events by using a kinematic variable
\begin{align}
\alpha_T=\frac{p_T^{j_2}}{M_{jj}}
\end{align}
which was presented to study SUSY signatures of similar topology in the final state \cite{Randall:2008rw}. 
The $p_T^{j_2}$ is the transverse momenta of the second leading jet while the $M_{jj}$ is the 
invariant mass of the dijet pair. The SM background trails off at 0.5 for back-to-back jets in QCD 
events and thus can help in looking for new physics signals which give large $\slashed{E}_T$ in the 
final state. This analysis has also been extended to final states with more than 2 jets \cite{Ozturk:2009fj}.
In addition one might require more specific set of selection cuts to isolate signals for new physics 
in final states with only jets and large missing transverse momenta \cite{Alves:2011sq}.  These techniques 
would also be useful in isolating our signal with $2j+\slashed{E}_T$ and $4j+\slashed{E}_T$ from the 
SM background.
\subsubsection*{\bf{1 lepton, 3 hard jets and $\slashed{E}_T$}}
The signal shown in Table \ref{tab:collider} is for LHC running with the center-of-mass energy of 
14 TeV. Without kinematic cuts the signal looks very promising. However, one expects several SM 
processes to give a similar final state. The most likely backgrounds to this process are $t \bar{t}, 
~tW,~t\bar{t}Z,~WZ$ and  $Wjj$ where one of the top quark decays semileptonically for the $t \bar{t}$ 
events, while either the top or $W$ gives the hard lepton for the $tW$ process, whereas the $W$ 
gives the lepton for the $WZ$ and $Wjj$ process. The backgrounds coming from the top quark 
processes can be suppressed using a $b$-jet veto on the signal, as our signal will only consist of the 
light quark jets.  As we demand that the final state has at least 3 hard jets, the $WZ$ and $Wjj$ will 
have to have one additional jet from radiations which should suppress this background significantly.
The single top SM background cross section is weak process and the cross section should be smaller 
and the strong selection cuts on the lepton, jets and the missing transverse energy should be enough 
to suppress this background. Similar weak process backgrounds like the triple gauge boson 
production should also give much smaller rates compared to the signal.
\subsubsection*{\bf{2 leptons, 2 hard jets and $\slashed{E}_T$}}
This process has a large $\sigma \times BR \simeq 500$ fb for $R^{-1}=400$ GeV at the current 
LHC energy.  In addition the signal should be decently clean with 2 hard leptons in the final state. 
The potentially dominant SM background comes from $t\bar{t}$ process with some contributions
from  $t \bar{t}Z$ and other weak processes like the $WWW,~WWZ$.  The background from pair 
production of tops will be very large at both $\sqrt{s}=7$ TeV and $\sqrt{s}=14$ TeV center-of-mass 
energy at LHC but can be again suppressed by using a $b$-jet veto on the signal.   In our model we 
have hard jets that are equally likely to be u, c, s, d, or b. This signal is promising but a full 
analysis of the SM background needs to be done to know whether the signal can stand out over 
the background.
\subsubsection*{\bf{4 leptons, 2 hard jets and $\slashed{E}_T$}}
The 4-lepton signal is very clean and promising as very few SM processes can lead to such 
final states with appreciable cross section. However, we do find a  low $\sigma \times BR = 30$ fb 
when compared to the other final states listed in Table \ref{tab:collider}.  With the available mass 
splitting for $R^{-1}=400$ GeV,  we expect the leptons to have at least $p_T>30$ GeV.  With 4 such 
hard leptons, 2 hard jets and substantial $\slashed{E}_T$, this signal should stand out over
the negligible SM background. The most likely sources for the SM background are again $t\bar{t}$ 
and triple or four vector boson processes, where the additional leptons in the $t\bar{t}$ process 
coming from the decays of the b-quarks. The strong $p_T$ cut on the leptons and the additional 
requirement of 2 hard jets should make these background completely negligible. With high enough 
luminosity, even higher values of $R^{-1}$ could be probed through this signal.

The multi-lepton signals with large jet multiplicity also leads to interesting correlations in the 
signature space with that of the parameters of the theory for different new physics models beyond 
the SM and can be a useful way of distinguishing signatures in our model with other models with 
similar signatures \cite{Bhattacherjee:2009jh}.
\subsubsection*{\bf{Monojet and $\slashed{E}_T$}}
In our model this signal is dominantly produced in the process, $pp\rightarrow N'N'g$.  It is a signal 
with a single jet plus missing $E_T$ with balancing transverse momenta, at the LHC. 
The $\sigma \times BR$ for this mode is around 2 pb where the jet has a minimum  $p_T$ cut of 20 GeV.  
The dominant SM background here comes from $Z+j$ in the hard $E_T$ regime where the Z decays 
invisibly while in the low $E_T$ regime it is dominated by QCD with mis-measurement of 
jets \cite{Vacavant:2001sd,Rizzo:2008fp}. As the single jet recoils against a massive system consisting of 
two heavy stable particles, it is expected to carry away a large transverse momenta which is balanced 
by the missing $E_T$. Thus a very high $\slashed{E}_T$ cut would make the signal significant against the
large SM background dominated by the $Z+j$ process. Another interesting signature would be the 
single photon signature much similar to the monojet signal, with smaller signal rates.
\subsubsection*{\bf{A very long-lived colored charged particle}}
A very interesting prediction in our model is the presence of a long-lived charged colored exotic 
fermion ($U^\prime$).  This particle is not listed in Table \ref{tab:collider} with the branching 
probabilities of the various exotics. Because of the parity assignment of the particle content in our
model, we find that the $U^\prime$ becomes very long lived and does not decay within the detector. 
Since it has to decay through the $A^\prime$ or the $A4_5^{\hat{a}}$ and the right-handed neutrino, 
it can only decay to a 5-body final state allowed by the kinematic phase space. The heavy right-handed 
neutrino gets a Majorana mass at the TeV scale through the VEV of the singlet scalar $S$ which breaks
the additional U(1) symmetry. Assuming a TeV-seesaw mechanism to be responsible for the light 
neutrino masses in the SM, the Dirac Yukawa couplings are of the order of $10^{-6}$. This would also 
be the strength of the mixing angle between the heavy singlet right-handed neutrino
and the light neutrinos. Thus the $U_R^\prime$ decays for example through the following decay chain:
$$ U_R^\prime \to A^{\prime *} N_R^* \to (u N^\prime) (e (W^*\to u\bar{d}))$$
Thus summing over all possible 5-body decays, we find that the width 
$\Gamma(U^\prime) \sim 10^{-27}$ GeV which gives a rough lifetime of 100 seconds. Thus it 
cannot decay within the detector. However, being a colored particle it would hadronize quickly to 
form charged or neutral hadrons that would interact and pass through the detector \cite{Kraan:2004tz}. 
Such scenarios can also happen in supersymmetric theories which have long-lived squarks, 
long-lived gluinos \cite{Giudice:2004tc, ArkaniHamed:2004yi} which form these type of hadrons. 
The neutral hadron would usually pass through the detector undetected, while the charged hadron would be slow moving highly ionizing particle leaving a charged track in the muon detector and 
passing through \cite{Drees:1990yw, Fairbairn:2006gg}. Thus one can have 1 or 2 heavily  
ionizing charged tracks in the detector passing through the muon chamber. Being a colored particle,  
the pair production cross section for the $U_R^\prime$ is quite large, as shown in 
Fig. \ref{fig:cross_sections}. Such a  signature will be a unique test of our model, complemented with 
the other signals listed above and also distinguishes our model from all other beyond SM theories. There
are already strong constraints on such particles from the LHC 
experiments \cite{Khachatryan:2011ts, Aad:2011hz}.

\section{A Five-Dimensional Supersymmetric Model}


In this Section, we shall also construct the five-dimensional 
 $N=1$ supersymmetric $SU(4)_C \times SU(2)_L \times U(1)_{I3R}$ models 
on $S^1/(Z_2\times Z'_2)$. 
The $N=1$ supersymmetric theory in five dimensions have 8 real
supercharges, corresponding to $N=2$ supersymmetry in four dimensions.  In
terms of the physical degrees of freedom, the vector multiplet
contains a vector boson $A_M$ with $M=0, 1, 2, 3, 5$, two Weyl
gauginos $\lambda_{1,2}$, and a real scalar $\sigma$.  In the
four-dimensional $N=1$ supersymmetry language, it contains a vector
multiplet $V \equiv (A_{\mu}, \lambda_1)$ and a chiral multiplet
$\Sigma \equiv ((\sigma+iA_5)/\sqrt 2, \lambda_2)$ which transform in
the adjoint representation of group $G$.  The five-dimensional
hypermultiplet consists of two complex scalars $\phi$ and $\phi^c$,
and a Dirac fermion $\Psi$.  It can be decomposed into two chiral
multiplets $\Phi(\phi, \psi \equiv \Psi_R)$ and $\Phi^c(\phi^c,
\psi^c \equiv \Psi_L)$, which are in the conjugate representations of
each other under the gauge group.

The general action for the group $G$ gauge fields and their
couplings to the bulk hypermultiplet $\Phi$ is~\cite{NAHGW}
\begin{eqnarray}
S&=&\int{d^5x}\frac{1}{k g^2}
{\rm Tr}\left[\frac{1}{4}\int{d^2\theta} \left(W^\alpha W_\alpha+{\rm H.
C.}\right)
\right.\nonumber\\&&\left.
+\int{d^4\theta}\left((\sqrt{2}\partial_5+ {\bar \Sigma })
e^{-V}(-\sqrt{2}\partial_5+\Sigma )e^V+
\partial_5 e^{-V}\partial_5 e^V\right)\right]
\nonumber\\&&
+\int{d^5x} \left[ \int{d^4\theta} \left( {\Phi}^c e^V {\bar \Phi}^c +
 {\bar \Phi} e^{-V} \Phi \right)
\right.\nonumber\\&&\left.
+ \int{d^2\theta} \left( {\Phi}^c (\partial_5 -{1\over {\sqrt 2}} \Sigma)
\Phi + {\rm H. C.}
\right)\right]~.~\,
\end{eqnarray}

Under the parity operator $P$, the vector multiplet transforms as
\begin{eqnarray}
V(x^{\mu},y)&\to  V(x^{\mu},-y) = P V(x^{\mu}, y) P^{-1}
~,~\,
\end{eqnarray}
\begin{eqnarray}
 \Sigma(x^{\mu},y) &\to\Sigma(x^{\mu},-y) = - P \Sigma(x^{\mu}, y) P^{-1}
~.~\,
\end{eqnarray}
For the hypermultiplet $\Phi$ and $\Phi^c$, we have~\cite{Li:2001tx}
\begin{eqnarray}
\Phi(x^{\mu},y)&\to \Phi(x^{\mu}, -y)  = \eta_{\Phi} P^{l_\Phi} \Phi(x^{\mu},y)
(P^{-1})^{m_\Phi}~,~\,
\end{eqnarray}
\begin{eqnarray}
\Phi^c(x^{\mu},y) &\to \Phi^c(x^{\mu}, -y)  = -\eta_{\Phi} P^{l_\Phi}
\Phi^c(x^{\mu},y) (P^{-1})^{m_\Phi} 
~,~\,
\end{eqnarray}
where $\eta_{\Phi}$ is $\pm$, $l_{\Phi}$ and $m_{\Phi}$ are
respectively the numbers of the fundamental index and anti-fundamental
index for the bulk multiplet $\Phi$ under the bulk gauge group $G$.
For example, if $G$ is an $SU(N)$ group, for fundamental
representation, $l_{\Phi}=1$, $m_{\Phi}=0$, and for adjoint
representation, $l_{\Phi}=1$, $m_{\Phi}=1$.  Moreover, the
transformation properties for the vector multiplet and hypermultiplets
under $P'$ are the same as those under $P$.

We consider the $SU(4)_C\times SU(2)_L\times U(1)_{I3R}$ models.
Let us explain our convention first. We denote the SM
quark doublets, right-handed up-type quarks, right-handed 
down-type quarks, lepton doublets, right-handed 
charged leptons, and right-handed neutrinos as
$Q_i$, $U_i^c$, $D_i^c$, $L_i$, $E_i^c$, and $N_i^c$,
respectively.  Similar to the non-supersymmetric models,
we introduce the matter fields
$FL_i$,  $\overline{FU}_i$, and $\overline{FD}_i$ with
the following quantum numbers under 
$SU(4)_C\times SU(2)_L\times U(1)_{I3R}$
gauge symmetry
\begin{eqnarray}
FL_i~=~{\mathbf{(4, 2, 0)}}~,~~ 
\overline{FU}_i ~=~{\mathbf{(\overline{4}, 1, -1/2)}}~,~~ 
\overline{FD}_i ~=~{\mathbf{(\overline{4}, 1, 1/2)}}~.~\,
\end{eqnarray}
The particle contents in $FL_i$, $FL_i^c$, $\overline{FU}_i$,
$\overline{FU}^c_i$, $\overline{FU}_i$, and $\overline{FU}^c_i$ are
\begin{align*}
& FL_j = (Q_j, L'_j), &   & FL_j^c = ((Q_j)^c, (L'_j)^c),  &
& FL_k = (Q'_{k-3}, L_{k-3}), \\ 
&  FL^c_k = ((Q^{\prime}_{k-3})^c, (L_{k-3})^c), & 
& \overline{FU}_j = (U^c_j, N^{\prime c}_j), &  &
\overline{FU}^c_j = ((U^{c}_j)^c , (N^{\prime c}_j)^c),  \\
& \overline{FU}_k = (U^{\prime c}_{k-3}, N^c_{k-3}), &   
&\overline{FU}^c_k = ( (U^{\prime c}_{k-3})^c, (N^c_{k-3})^c), & 
& \overline{FD}_j = (D^c_j, E^{\prime c}_j),  \\  
& \overline{FD}^c_j = ((D^{c}_j)^c , (E^{\prime c}_j)^c), &
& \overline{FD}_k = (D^{\prime c}_{k-3}, E^c_{k-3}), &  
& \overline{FD}^c_k = ( (D^{\prime c}_{k-3})^c, (E^c_{k-3})^c),
\end{align*}
where $j=1,~2,~3$, and $k=4,~5,~6$. Thus,
$FL_i$ contain the left-handed quarks and leptons,
$\overline{FU}_i$ contain the right-handed up-type quarks
and neutrinos, and $\overline{FD}_i$ contain the
right-handed down-type quarks and charged leptons.

We will break the $SU(4)_C$ gauge symmetry down to
the $SU(3)_C\times U(1)_{B-L}$ gauge symmetry
by orbifold projections.
To break the $U(1)_{B-L}\times U(1)_{I3R}$ gauge symmetries
down to the $U(1)_Y$ gauge symmetry,
we introduce one pair of SM singlet Higgs fields
$S$ and $\overline{S}$
on the 3-brane at $y=\pi R/2$ where only the
$SU(3)_C\times SU(2)_L\times U(1)_{B-L} \times U(1)_{I3R}$
gauge symmetries
are preserved. The quantum numbers for $S$ and $\overline{S}$
are ${\mathbf{(1, 1, -1, 1)}}$ and 
${\mathbf{(1, 1, 1, -1)}}$, respectively.
To break the electroweak gauge symmetry, we introduce
one pair of Higgs doublets $H_u$ and $H_d$ in the bulk
whose quantum numbers under 
$SU(4)_C\times SU(2)_L\times U(1)_{I3R}$ are
\begin{eqnarray}
H_u~=~{\mathbf{(1, 2, {1\over 2})}}~,~~ 
H_d~=~{\mathbf{(1, 2, - {1\over 2})}}~.~\,
\end{eqnarray}

Although the gauge symmetry breaking is similar to the
non-supersymmetric model, we will still discuss it
since we also need to break the five-dimensional $N=1$ 
supersymmetry down to the four-dimensional $N=1$ supersymmetry.
To break the $SU(4)_C$ gauge symmetry down to the
$SU(3)_C\times U(1)_{B-L}$ gauge symmetry, and
to  project out the zero modes of
the fifth component of gauge fields, and
$FL_i^c$, $\overline{FU}_i^c$, and  $\overline{FD}^c_i$,
we choose the following $4\times 4$ matrix representations 
for the parity operators $P$ and $P'$
\begin{equation}
\label{eq:pppsu5}
P={\rm diag}(+1, +1, +1, +1)~,~P'={\rm diag}(+1, +1, +1, -1)
 ~.~\,
\end{equation}
Under the $P$ parity,  the five-dimensional $N=1$ 
supersymmetry is broken down to the four-dimensional $N=1$ supersymmetry.
And under the $P'$ parity, the $SU(4)_C$ gauge symmetry is broken
down to the  $SU(3)_C\times U(1)_{B-L}$ gauge symmetry. Thus,
the five-dimensional $N=1$ supersymmetric $SU(4)_C$ gauge symmetry is
broken down to the four-dimensional $N=1$ supersymmetric 
$SU(3)_C\times U(1)_{B-L}$ gauge
symmetry for the zero modes. For the zero modes and KK modes, the
four-dimensional $N=1$ supersymmetry is preserved on the 3-branes at the
fixed points, and only the $SU(3)_C\times U(1)_{B-L}$ 
 gauge symmetry is preserved on the
3-brane at $y=\pi R/2$~\cite{Li:2001tx}.

By the way, for the $SU(2)_L$ gauge symmetry, we choose
\begin{equation}
P~=~P'~=~{\rm diag}(+1, +1)
 ~.~\,
\end{equation}
And for the $U(1)_{I3R}$ gauge symmetry, we choose $P=P'=1$.
Thus, the  five-dimensional $N=1$ supersymmetric 
$SU(2)_L\times U(1)_{I3R}$ gauge symmetry is
broken down to the four-dimensional $N=1$ supersymmetric 
 $SU(2)_L\times U(1)_{I3R}$ gauge
symmetry for the zero modes. 

We denote the vector multiplets for $SU(4)_C$, $SU(2)_L$,
and $U(1)_{I3R}$ as $(V4_{\mu}^A , \Sigma4^A)$,
$(V2_{\mu}^A , \Sigma2^A)$, and $(V1_{\mu} , \Sigma1)$,
respectively. In addition, for $FL_i$, $\overline{FU}_i$,
and $\overline{FD}_i$, we choose $\eta=+1$ and $\eta'=+1$
for $i=1,~2,~3$, and choose  $\eta=+1$ and $\eta'=-1$
for $i=4,~5,~6$. 
For $H_u$ and $H_d$, we choose  $\eta=+1$ and $\eta'=+1$.
The particle spectra and their $(P,~P')$ parity are given
in Table~\ref{Spectra-I}.

\begin{table}[htb]
\begin{center}
\begin{tabular}{|c|c|c|}
\hline
$(P,P')$ & ${\rm Field}$ & ${\rm Mass}$ \\ \hline
$(+,+)$ & $V4_{\mu}^a$, $V2_{\mu}^A$, $V1_{\mu}$,
$Q_i$, $U_i^c$, $D_i^c$, $L_i$, $N_i^c$, $E_i^c$, $H_u$, $H_d$
  & ${{2n}\over R}$ \\ \hline
$(+,-)$ &  $V4_{\mu}^{\hat{a}}$, 
$Q'_i$, $U_i^{\prime c}$, $D_i^{\prime c}$, $L'_i$, $N_i^{\prime c}$, $E_i^{\prime c}$
  & ${{2n+1}\over R}$  \\ \hline
$(-,+)$ &  $\Sigma4^{\hat{a}}$, 
$(Q'_i)^c$, $(U_i^{\prime c})^c$, $(D_i^{\prime c})^c$, 
$(L'_i)^c$, $(N_i^{\prime c})^c$, $(E_i^{\prime c})^c$
  & ${{2n+1}\over R}$  \\ \hline
$(-,-)$ & ~$\Sigma4^a$, $\Sigma2^A$, 
$\Sigma1$, 
$(Q_i)^c$, $(U_i^c)^c$, $(D_i^c)^c$, $(L_i)^c$, $(N_i^c)^c$, $(E_i^c)^c$, 
$H^c_u$, $H^c_d$~
  & ${{2n+2}\over R}$ \\ \hline
\end{tabular}
\end{center}
\caption{Parity assignments and masses ($n\ge 0$) for  the bulk fields. }
\label{Spectra-I}
\end{table}

In our models, the Yukawa couplings in the superpotential are
\begin{eqnarray}
W &=& \left(h^u_{ij} FL_{i} \overline{FU}_j H_u + h^d_{ij}
FL^i \overline{FD}_j H_d + h^{\prime u}_{ij} FL^{c}_{i} 
\overline{FU}^c_j H_u + h^{\prime d}_{ij}
FL^c_i \overline{FD}^c_j H_d  \right) \times \delta(y)  \nonumber \\ 
&&
+ \left(y^u_{ij} Q_i U^c_j H_u + y^d_{ij} Q_i D^c_j H_d
+ y^{\nu}_{ij} L_i N_j^c H_u + y^e_{ij} L_i E_j^c H_d
 + y^N_{ij} S  N_i^c N_j^c \right) \nonumber \\ 
&& \times \delta(y-\pi R/2)~,~\,
\end{eqnarray}
where $h^u_{ij}$, $h^d_{ij}$, $h^{\prime u}_{ij}$ and 
$h^{\prime d}_{ij}$ are non-zero  only for  $1\le i,~j \le 3$ and 
 $4\le i,~j \le 6$.
Because $S$ will obtain a VEV around 1 TeV, the active
neutrinos will obtain the observed masses via seesaw
mechanism if $h^u_{ij}$
are about $10^{-5}$ for $4\le i,~j \le 6$.

Next, let us discuss the $U(1)_{B-L}\times U(1)_{I3R}$ gauge symmetry 
breaking. We denote the gauge fields for $U(1)_{B-L}$ and
$U(1)_Y$ as $A^{B-L}_{\mu}$ and $A^R_{\mu}$, respectively.
To give the VEVs to $S$ and $\overline{S}$, we consider
the following superpotential
\begin{eqnarray}
W &=& S' \left(S \overline{S} -M_S^2 \right) ~,~\,
\end{eqnarray}
where $S'$ is the SM singlet fields, and $M_S$ is a mass parameter.
From the F-term flatness for $S'$, we obtain that
$S$ and $\overline{S}$ will respectively acquire VEVs $v'_1$ and $v'_2$
where $v_1' v_2' \simeq M_S^2$. Thus, 
the $U(1)_{B-L}\times U(1)_{I3R}$ gauge symmetries are broken
down to the $U(1)_Y$ gauge symmetry by Higgs mechanism. 


Furthermore, let us comment on the dark matter in our models.
There are  $Z_2\times Z'_2$ symmetries in our models, thus,
 we can have two dark matter candidates. Including the
radiative corrections, we find that one dark matter candidate
from one linear combination of $N_i^{\prime c}$ and the neutral
components of $L'_i$,
and the other dark matter candidate
from one linear combination of $(N_i^{\prime c})^c$ and the neutral
components of $(L'_i)^c$ after we consider the Yukawa contributions.
At the LHC, $A_{\mu}'$ and $\overline{A'}_{\mu}$ can be produced
via gluon fusions, and then decays.
The interesting LHC signals in our models are the di-leptons
and four leptons plus jets for final states. Thus, we may want 
 that $Q'_i$ are lighter than $A_{\mu}'$ and 
$\overline{A'}_{\mu}$, which can be realized
via the Yukawa interactions. 

For simplicity, we consider $h_{ij}^u$ and $h_{ij}^{\prime u}$
are diagonal for $4 \le i,~j \le 6$. Thus, the mass
matrix for $(Q'_i, ~(Q'_i)^c, ~U_i^{\prime c}, ~(U_i^{\prime c})^c) $ is
\begin{eqnarray}
M=
\left(
\begin{array}{cccc}
0 &  M_i & m_i & 0 \\
M_i & 0 & 0 & m'_i \\
m_i & 0 & 0 & M_i \\
0 & m'_i & M_i & 0
\end{array}
\right)
~,~\,
\end{eqnarray}
where $M_i$ is $1/R$ plus the radiative corrections,
and 
\begin{eqnarray}
m_i~=~ h_{(i+3) (i+3)}^u ~\langle H_u \rangle~,~~
m'_i~=~ h_{(i+3) (i+3)}^{\prime u} ~\langle H_u \rangle  
~.~\,
\end{eqnarray}
Because $m_i$ are also parts of the  neutrino Dirac masses,
we assume they are very small about $10^{-3}$ GeV or smaller. 
Otherwise, we have to fine-tune the neutrino Dirac masses 
on the 3-branes at
two  fixed points so that the neutrino Dirac masses
are about $10^{-3}$ GeV or smaller.
Thus, choosing $m_i=0$ we obtain the following four
eigenvalues
\begin{eqnarray}
\lambda_{1,2} ~=~ \pm {1\over 2} \left(m'_i -{\sqrt \Delta}\right)
~,~~\lambda_{3,4} ~=~ \pm {1\over 2} \left(m'_i +{\sqrt \Delta}\right)
~,~\,
\end{eqnarray}
where
\begin{eqnarray}
\Delta ~=~ 4M_i^2+ m^{\prime 2}_i~.~\,
\end{eqnarray}
And the corresponding eigenvectors are
\begin{eqnarray}
V_{1,2} &=& \left( -{{m_i'+{\sqrt \Delta}}\over {2M_i}}, \pm 1,
\mp {{2M_i}\over {-m_i'+{\sqrt \Delta}}}, 1 \right)~,~\, \nonumber \\
V_{3,4} &=& \left( -{{m_i'-{\sqrt \Delta}}\over {2M_i}}, \pm 1,
\pm {{2M_i}\over {m_i'+{\sqrt \Delta}}}, 1 \right)~.~\,
\end{eqnarray}
Thus, we can indeed have two relatively light states 
and two relatively heavy states with large
mixings. Similarly, we can study the mass matrix for
$(L'_i, ~(L'_i)^c, ~N_i^{\prime c}, ~(N_i^{\prime c})^c) $, and get the
corresponding eigenvalues and eigenvectors. In this
case, the two light neutral states are dark matter candidates.


\section{Summary and Conclusions}

Motivated by possible new physics which can be observed at the LHC,
we have proposed  a model which have leptoquarks as color triplet
gauge bosons at the TeV scale. Such color triplet gauge boson, though
exist in the grand unified theories, such as $SU(5)$ or $SO(10)$, they 
cause proton decay, and their masses are of the order of $10^{16}$ GeV.
In the partially unified models of Pati-Salam type, these leptoquark 
gauge bosons do not cause proton decay. However, they can cause rare
meson decays which set the lower bound on their masses to be greater than
$2,300$ TeV. This motivated us to construct a model of partial unification
based on the gauge symmetry $SU(4)_C \times SU(2)_L \times U(1)_{I3R}$ in
{\it five} dimensions. This symmetry is broken to the usual SM gauge symmetry
in four dimensions by compactification on an orbifold at a TeV scale. 
In order to avoid the constraint from the rare meson decays, we paired 
up SM quarks with new leptons, and the SM leptons with new quarks.  These
new quarks and leptons in the model have no zero modes, have only their KK 
excitations in the spectrum with masses in the TeV. These KK excitations are also vector-like,
and thus avoid any problem with the precision electroweak parameters. Furthermore, 
since the leptoquark gauge bosons in the model couple only to quarks and new leptons, or leptons and new quarks, their high mass limits from the rare meson decays are
also avoided.

Because they are colored, these leptoquark gauge bosons, as well as the
new quarks with masses in the TeV scale can be produced at the LHC with 
large cross sections.  The lightest particle in the spectrum is a neutral lepton,
and is stable, giving the possibility of being a candidate for the dark matter. We have calculated the
mass spectrum and the decay modes of these new particles in the model.
The final state signals are high $p_T$ jets and high $p_T$ leptons with large 
missing energy.  The most dominant signals are 2 charged leptons and 2 hard jets plus 
missing $E_T$ or 1 charged lepton and 3 hard jets plus missing transverse 
energy, and are observable at the LHC. The four lepton signal with at lease 2 hard jets, 
though small, will have very little  SM background, and should be observable. 
The model also has a large monojet signal with a high $p_T$ jet and large missing
energy. Finally, one of the new quarks (up-type) has a five-body decay mode, and is long-lived.
It will hadronize, and these charged hadrons will leave  tracks as they pass through the detector.

\begin{acknowledgments}


We are very grateful to many helpful discussions with J. Lykken. We also thank 
G. Giudice, D. Hooper and G. Servant for several useful discussions. SN was a summer visitor at Fermilab and CERN when this work was in progress, and he thanks the Fermilab Theory
Group and CERN Theory Group for warm hospitality and support during these visits. This research was supported in part by the Natural Science Foundation of China under grant numbers 10821504 and 11075194 (TL), by the United States Department of Energy Grant Numbers DE-FG03-95-Er-40917, DE-FG02-04ER41306 and DE-FG02-04ER46140.

\end{acknowledgments}


\end{document}